\documentclass[twocolumn,showkeys,preprintnumbers,amsmath,amssymb]{revtex4}

\usepackage{graphicx}
\usepackage{dcolumn}
\usepackage{bm}

\def\kmsmpc{${\rm ~km ~s^{-1}Mpc^{-1}}$}

\begin{document}


\title{Spiral galaxy rotation curves determined from Carmelian general relativity}

\author{John G. Hartnett}
 \email{john@physics.uwa.edu.au}
\affiliation{School of Physics, the University of Western Australia\\35 Stirling Hwy, Crawley 6009 WA Australia}%

\date{\today}

\begin{abstract}
Equations of motion, in cylindrical co-ordinates, for the observed rotation of gases within the gravitational potential of spiral galaxies have been derived from Carmeli's Cosmological General Relativity theory. A Tully-Fisher type relation results and rotation curves are reproduced without the need for non-baryonic halo dark matter. Two acceleration regimes are discovered that are separated by a critical acceleration $\approx 4.75 \times 10^{-10}$ $m.s^{-2}$. For accelerations larger than the critical value the Newtonian force law applies, but for accelerations less than the critical value the Carmelian regime applies. In the Newtonian regime the accelerations fall off as $r^{-2}$, but in the Carmelian regime the accelerations fall off as $r^{-1}$. This is new physics but is exactly what is suggested by Milgrom's phenomenological MOND theory.

\end{abstract}

\keywords{Carmeli's Cosmological General Relativity, Tully-Fisher, galaxy rotation curves}
\maketitle

\section{\label{sec:Intro}Introduction}
The rotation curves highlighted by the circular motion of stars or more accurately characterised by the spectroscopic detection of the motion of neutral hydrogen and other gases in the disk regions of spiral galaxies have caused concern for astronomers for many decades. Newton's law of gravitation predicts much lower orbital speeds than those measured in the disk regions of spiral galaxies. 

The most luminous galaxies show slightly declining rotation curves (orbital speed vs radial position from nucleus) in the regions outside the star bearing disk, coming down from a broad maximum in the disk. Intermediate mass galaxies have mostly nearly flat rotation speeds along the disk radius. Lower luminosity galaxies usually have monotonically increasing orbital velocities across the disk. 

The traditional solution has been to invoke halo `dark matter'   \cite{Begeman1991} that surrounds the galaxy but is transparent to all forms of electromagnetic radiation. In fact, astronomers have traditionally resorted to `dark matter' whenever known laws of physics were unable to explain the observed dynamics.

In 1983 Milgrom introduced his MOND \cite{Milgrom1983a,Milgrom1983b,Milgrom1983c}, an empirical approach, which attempts to modify Newtonian dynamics in the region of very low acceleration. Newton's law describes a force proportional to $r^{-2}$, where $r$ is the radial position from the center of the matter distribution, but Milgrom finds that a $r^{-1}$ law fits the data very well \cite{Begeman1991}. 

Carmeli \cite{Carmeli2000,Carmeli2002} approached the problem from a different perspective. He formulated a modification, actually an extension of Einstein's general theory, in an expanding universe taking into account the Hubble expansion as a fundamental axiom, which imposes an additional constraint on the dynamics of particles \cite{Carmeli1982}. 

Carmeli believes the usual assumptions in deriving Newton's gravitational force law from general relativity are insufficient, that gases and stars in the arms of spiral galaxies are not immune from Hubble flow. As a consequence a universal constant $a_{0}$ (in this case, slightly different to Milgrom's) is introduced as a characteristic acceleration in the cosmos.  Using this theory Carmeli successfully provided a theoretical description of the Tully-Fisher law \cite{Carmeli1998}. 

Following Carmeli's lead, Hartnett \cite{Hartnett2005} showed that the same line of reasoning leads to plausible galaxy rotation curves. The latter used a density model for spiral galaxies, that assumed that most of the mass of the galaxy was in the nuclear bulge and that the density in the disk region was constant. However by perturbing the density one could also get the variation in rotation speeds as typically observed. That paper also assumed spherical coordinates and a hyperbolic density distribution of disk matter, conditions which are not appropriate for exponential-density-model galaxies. 

In this paper we take the analysis further, and in a more rigorous way, model the gravitational potential and the resulting forces determining how test particles move in the disks of spiral galaxies using cylindrical coordinates and an exponential density distribution. Two acceleration regimes are discovered. In one, normal Newtonian gravitation applies. In that regime the effect of the Hubble expansion is not observed or is extremely weak. It is as if the particles' accelerations are are so great that they slip across the expanding space. In the other, new physics is needed. There the Carmelian metric provides it. In this regime the accelerations of particles are so weak that their motions are dominated by the Hubble expansion and as a result particles move under the combined effect of both the Newtonian force and a post-Newtonian contribution. 

\section{\label{sec:Gpotential}Gravitational potential}

In the weak gravitational limit, where Newtonian gravitation applies, it is sufficient to assume the Carmeli metric with non-zero elements $g_{00} = 1 + 2 \phi /c^{2}$, $g_{44} = 1 + 2 \psi /\tau^{2}$, $g_{kk} = -1$, ($k = 1, 2, 3$) in the lowest approximations in both $1/c$ and $1/\tau$. Here a new constant, called the Hubble-Carmeli constant, is introduced $\tau \approx 1/H_{0}$. See \cite{Carmeli2002} for details. The potential functions $\phi$ and $\psi$ are determined by Einstein's field equations and from their respective Poisson equations,
\begin{subequations}
\begin{eqnarray} 
\nabla^{2} \phi = 4 \pi G \rho_{eff} \label{eqn:phi},
\\
\nabla^{2} \psi = \frac{4 \pi G \rho_{eff}}{a_{0}^{2}} \label{eqn:psi},
\end{eqnarray}
\end{subequations}
\\
where  $\rho$ is the mass density and $a_{0}$ a universal characteristic acceleration $a_{0} =c/\tau$. As usual $c$ is the speed of light in vacuo. 

In Carmelian theory $\rho_{eff} = \rho -\rho_{c}$ is used instead of matter density $\rho$. The parameter $\rho_{c}$ is the critical density of the universe given by $\rho_{c} =3/8 \pi G \tau^{2} \approx 10^{-29} \; g.cm^{-3}$. However in a galaxy because $\rho \gg \rho_{c}$,  $\rho_{c}$ can be neglected in this paper.

A comparison of $\phi$ and $\psi$ in (\ref{eqn:phi}) and (\ref{eqn:psi}) leads to $\psi = \phi /a_{0}^{2}$ within an arbitrary additive constant. Since both potentials are defined with respect to the same co-ordinate system, in reality, we only need deal with one potential function, the gravitational potential, $\phi$.

In cylindrical coordinates ($r,\theta, z$) the potential $\phi$ that satisfies (\ref{eqn:phi}) can be found from \cite{Toomre1963},
\begin{equation} \label{eqn:potential}
\phi (r) = -2 \pi G \int^{\infty}_{0} J_{0}(k r)dk \int^{\infty}_{0} \rho(r') J_{0}(k r') r' dr' 
\end{equation}
where $J_{0}(kr)$ is the zeroth order Bessel function function and $k$ is the $z$ coordinate scale factor ($k = 1/b$). It is also assumed that the density function can be modeled as a delta function of the vertical coordinate $z$. Therefore the density $\rho(r,z) = \rho(r)\rho(z) = \rho(r)\delta(z)$ with no $\theta$ dependence. To correctly model the effect of the spiral arms a $\theta$ dependence may be needed, but for our model it is assumed independent. The requirement on the $z$ dependence is satisfied with density functions of the form 
\begin{equation} \label{eqn:zdensity}
\rho(z) = \frac{1}{2 b} sech(\frac{z}{b})^{2} \ \ or \ \ e^{-|z|/b}.
\end{equation}
Here, provided the scale length $b$ is much smaller than the limit of the actual matter distribution in the $z$ direction then the integral over all $z$ yields a contribution to the mass of unity. This is the thin disk approximation which seems to be fairly applicable over both disk and galactic bulge.

The integral over $dk$ in (\ref{eqn:potential}) is the surface density which may be calculated once the form of the density $\rho$ is known. Following from observation we choose an exponential function of the form
\begin{equation} \label{eqn:rdensity}
\rho(r) = \frac{M}{2 \pi a^2}e^{-r/a}.
\end{equation}
for the radial dependence, where $a$ is a radial scale length and $M$ is the mass of the galaxy.
\begin{figure}
\includegraphics[width = 3.5 in]{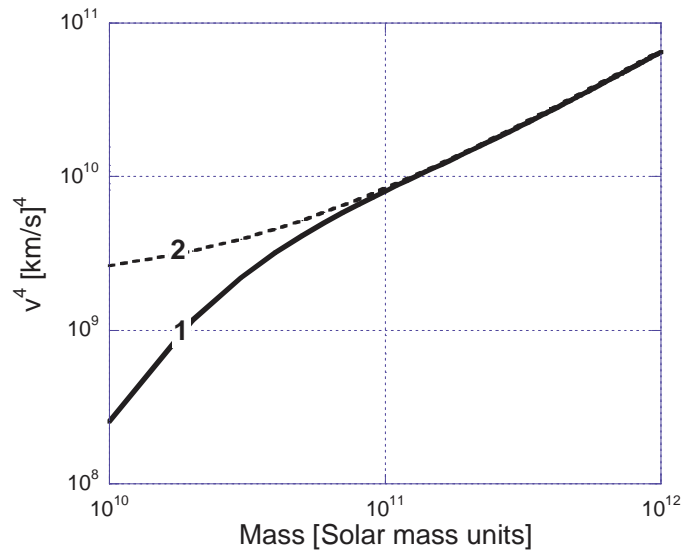}
\caption{\label{fig:fig1} Tully-Fisher law plotted on logarithmic axes. Curve 1 (solid line) represents the fourth order dependence of the rotational speeds of tracer gases in galaxies determined from the Carmelian equation (\ref{eqn:Carmelian}). The masses are expressed in solar mass units of $M_{\odot} = 2 \times 10^{30} kg$. Curve 2 (broken line) represents the straight line $v^4 = 2 \times 10^9 + 0.064 M$ (Mass) }
\end{figure}

\section{\label{sec:Eqnmotion}Equations of Motion}

The Hubble law describes the expansion of the cosmos and the matter embedded in it. Therefore the line element for any two points in this `new' \textit{space-time-velocity} is $ds^{2} = g_{00} c^{2} dt^{2}+ g_{kk} (dx^{k})^2+ g_{44} \tau^{2} dv^{2} = 0$. Here $k = 1, 2, 3$. The relative separation in 3 spatial coordinates $r^{2} = (x^{1})^{2} + (x^{2})^{2} + (x^{3})^{2}$ and the relative velocity between points connected by $ds$ is $v$. The Hubble-Carmeli constant,  $\tau$, is a constant for all observers at the same epoch, therefore may be regarded as a constant on the scale of any measurements.

The equations of motion (B.62a and B.63a from \cite{Carmeli2002}) to lowest approximation in $1/c$ are reproduced here,
\begin{equation} \label{eqn:derivphi}
\frac{d^{2} x^{k}} {d t^{2}} = - \frac{1}{2}\frac{\partial \phi} {\partial x^{k}}.
\end{equation}
This is the usual looking geodesic equation derived from general relativity but now in 5 dimensions. And the second is a new `phase space' equation derived from the Carmeli theory, \cite{Carmeli2002}
\begin{equation} \label{eqn:derivpsi}
\frac{d^{2} x^{k}} {d v^{2}} = - \frac{1}{2}\frac{\partial \psi} {\partial x^{k}}.
\end{equation}
\begin{figure}
\includegraphics[width = 3.5 in]{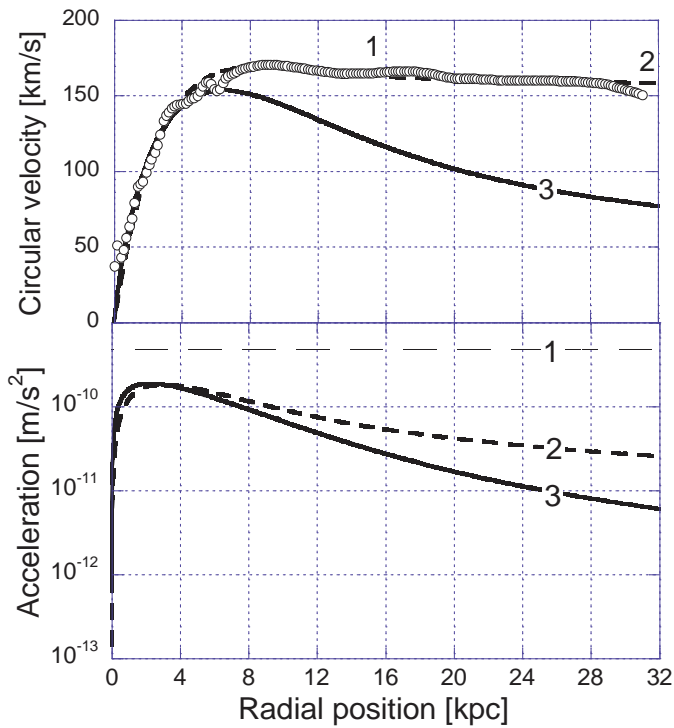}
\caption{\label{fig:fig2} (a) Above: The rotational speeds of tracer gases in NGC 3198 (SBc barbed spiral)(circles - curve 1). Theoretical curve fits from the Carmelian equation (\ref{eqn:rotcurve}) (broken curve 2) and from the Newtonian equation (\ref{eqn:Newtonian2}) (curve 3) (b) Below: The critical acceleration $\frac{2}{3}a_{0}$ (curve 1). The rotational accelerations determined from the Carmelian (curve 2) and the Newtonian (curve 3) equations with their respective values of $a$ and $M$}
\end{figure}
\begin{figure}
\includegraphics[width = 3.5 in]{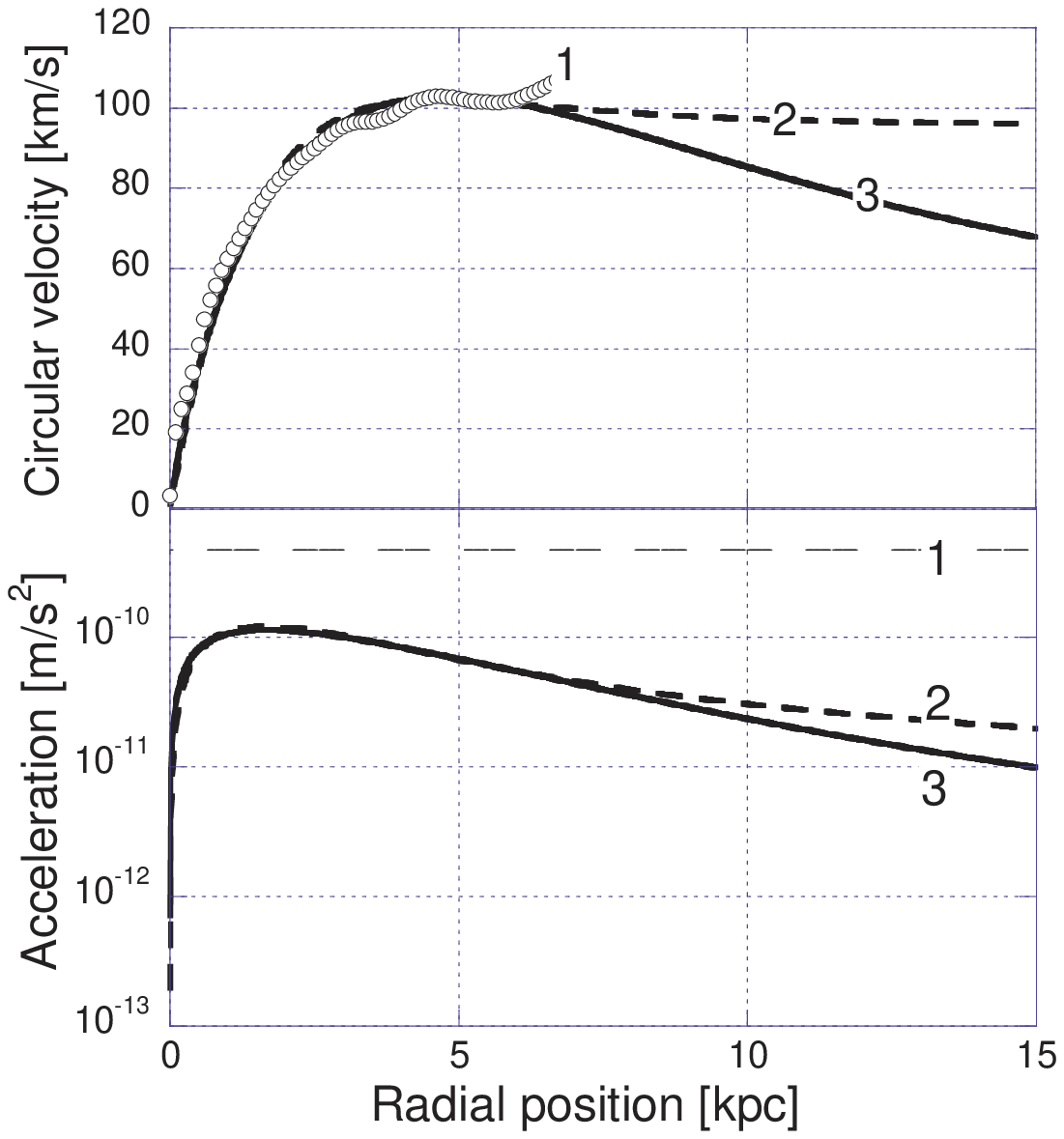}
\caption{\label{fig:fig3} (a) Above: The rotational speeds of tracer gases in NGC 0598 (Sc spiral) (circles - curve 1).  Theoretical curve fits from the Carmelian equation (\ref{eqn:rotcurve}) (curve 2) and from the Newtonian equation (\ref{eqn:Newtonian2}) (curve 3). (b) Below: The critical acceleration $\frac{2}{3}a_{0}$ (curve 1). The corresponding rotational accelerations determined from the Carmelian (curve 2) and the Newtonian (curve 3) equations}
\end{figure}

\subsection{\label{sec:Nequation}Newtonian}

It follows from (\ref{eqn:derivphi}), (\ref{eqn:rdensity}) and (\ref{eqn:potential}), and  the usual form of the circular motion equation 
\begin{equation} \label{eqn:Newtonian}
\frac{v^{2}}{r} = \frac{d \phi} {dr},
\end{equation}
that
\begin{equation} \label{eqn:Newtonian2}
v^{2} = \frac{G M r^{2}}{2 a^{3}} \Pi,
\end{equation}
where $G$ denotes the gravitational constant and 
\begin{equation} \label{eqn:Besselfn}
\Pi = I_{0}\left(\frac{r}{2 a}\right)K_{0}\left(\frac{r}{2 a}\right)- I_{1}\left(\frac{r}{2 a}\right)K_{1}\left(\frac{r}{2 a}\right)
\end{equation}
where $I$ and $K$  are standard zeroth and first order Bessel functions. 

\begin{figure}
\includegraphics[width = 3.5 in]{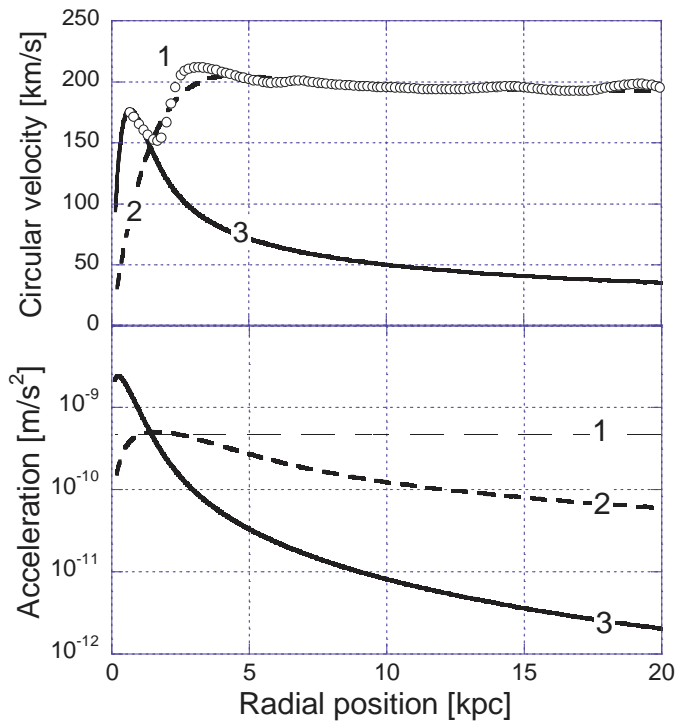}
\caption{\label{fig:fig4} (a) Above: The rotational speeds of tracer gases in NGC 2903 (Sc spiral) (circles - curve 1).  Theoretical curve fits from the Carmelian equation (\ref{eqn:rotcurve}) (curve 2) and from the Newtonian equation (\ref{eqn:Newtonian2}) (curve 3) (b) Below: The critical acceleration $\frac{2}{3}a_{0}$ (curve 1). The corresponding rotational accelerations determined from the Carmelian (curve 2) and the Newtonian (curve 3) equations}
\end{figure}

Equation (\ref{eqn:Newtonian2}) is the usual Newtonian result for the speed of circular motion in a cylindrical gravitational potential. This equation has been plotted in curve 3 of figs \ref{fig:fig2}(a) - \ref{fig:fig10}(a) as a function of radial position from the center of a galaxy in kiloparsecs ($kpc$) where $kpc \approx 3 \times 10^{19}$ m. Best fits were determined with $M$ and $a$ as free parameters. Through out this paper $M$ is expressed in solar mass units $M_{\odot} \approx 2 \times 10^{30}$ kg.

\subsection{\label{sec:Cequation}Carmelian}

Using $\psi = \phi /a_{0}^{2}$ in (\ref{eqn:derivpsi}) results in a new equation
\begin{equation} \label{eqn:Carmelieqn}
v = a_{0}\int^{r}_{0}\frac{dr}{\sqrt{-\phi}},
\end{equation}
where we have integrated and solved for $v$ as a function of $r$. Using the potential $\phi$, determined from (\ref{eqn:rdensity}) and (\ref{eqn:potential}), in (\ref{eqn:Carmelieqn}), results in 
\begin{equation} \label{eqn:Carmelieqngoodapprox}
v = \frac{2}{3}a_{0} \frac{r^{3/2}}{\sqrt{GM}},
\end{equation}
which describes the expansion of space within a galaxy. 

It must be realised that the only direction in the cylindrical coordinates of a galaxy that is free to expand in the Hubble flow is the azimuthal \cite{Hartnett2005}. Therefore (\ref{eqn:Carmelieqngoodapprox}) describes the velocity component in that direction. Carmeli applied this line of reasoning \cite{Carmeli1998,Carmeli2000}. 

To establish the combined result of the two equations of motion (\ref{eqn:Newtonian2}) and (\ref{eqn:Carmelieqngoodapprox}), the simultaneous speed of test particles must be determined by the elimination of $r$ between the two equations. The physical meaning can be understood in terms of particles that simultaneously satisfy both (\ref{eqn:derivphi}) and (\ref{eqn:derivpsi}). 

The usual Newtonian expression (\ref{eqn:derivphi}) describes motion under the central potential but assumes that spatial coordinates are fixed. Whereas the new equation (\ref{eqn:derivpsi}) describes the expansion of space itself within a galaxy. Therefore we must find the combined (simultaneous) effect of (\ref{eqn:Newtonian2}) and (\ref{eqn:Carmelieqngoodapprox}). The result is a post-Newtonian equation, 
\begin{equation} \label{eqn:Carmelian}
v^{2/3} = \frac{(G M)^{5/3} }{(\frac{2}{3} a_{0})^{4/3} 2 a^{3}} \Pi,
\end{equation}
derived from (\ref{eqn:Newtonian2}) where the following substitution
\begin{displaymath} 
r \rightarrow \left(\frac{G M v^2}{(\frac{2}{3} a_{0})^2} \right)^{1/3},
\end{displaymath}
has been made from (\ref{eqn:Carmelieqngoodapprox}).
The resulting equation, hereafter referred to as Carmelian, cannot be solved analytically. However using the Mathematica software package (\ref{eqn:Carmelian}) can be solved numerically. 

The result is plotted in curve 1 of fig. \ref{fig:fig1} where it has been assumed that $a = 1\; kpc$ and it is compared with the straight line $v^4 = 2 \times 10^9 + 0.064 M$ (curve 2). For large $M$ the small offset can be neglected. This result indicates that the fourth order dependence on rotational speed ($v$) is directly proportional to mass ($M$) for large masses. 

Assuming that the masses, of the galaxies studied, are directly proportional to their luminosity, this dependence then becomes the Tully-Fisher relationship. This extends the work of Carmeli \cite{Carmeli2000}, and derives the underlying theoretical framework upon which the Tully-Fisher law is founded. 

\subsection{\label{sec:Rotationcurve}Rotation curves}

In \cite{Carmeli1998, Carmeli2000, Hartnett2005}, using spherical co-ordinates, it was found that in the limit of large $r$ and where all the matter was interior to the position of a test particle, such a particle is also subject to an additional circular motion described by (\ref{eqn:Carmelieqngoodapprox}). Apparently this is the result of the expansion of space itself within the galaxy but in an azimuthal direction to the usual center of coordinates of the galaxy. In this paper also the same result (\ref{eqn:Carmelieqngoodapprox}) was obtained but in this case derived using cylindrical coordinates.
\begin{figure}
\includegraphics[width = 3.5 in]{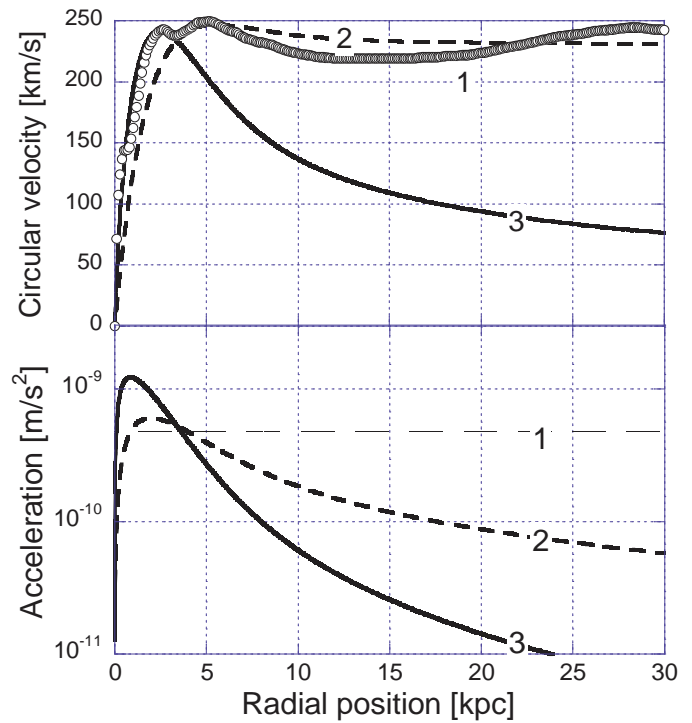}
\caption{\label{fig:fig5} (a) Above: The rotational speeds of tracer gases in NGC 7331 (Sbc spiral) (circles - curve 1).  Theoretical curve fits from the Carmelian equation (\ref{eqn:rotcurve}) (curve 2) and from the Newtonian equation (\ref{eqn:Newtonian2}) (curve 3) (b) Below: The critical acceleration $\frac{2}{3}a_{0}$ (curve 1). The corresponding rotational accelerations determined from the Carmelian (curve 2) and the Newtonian (curve 3) equations}
\end{figure}
\begin{figure}
\includegraphics[width = 3.5 in]{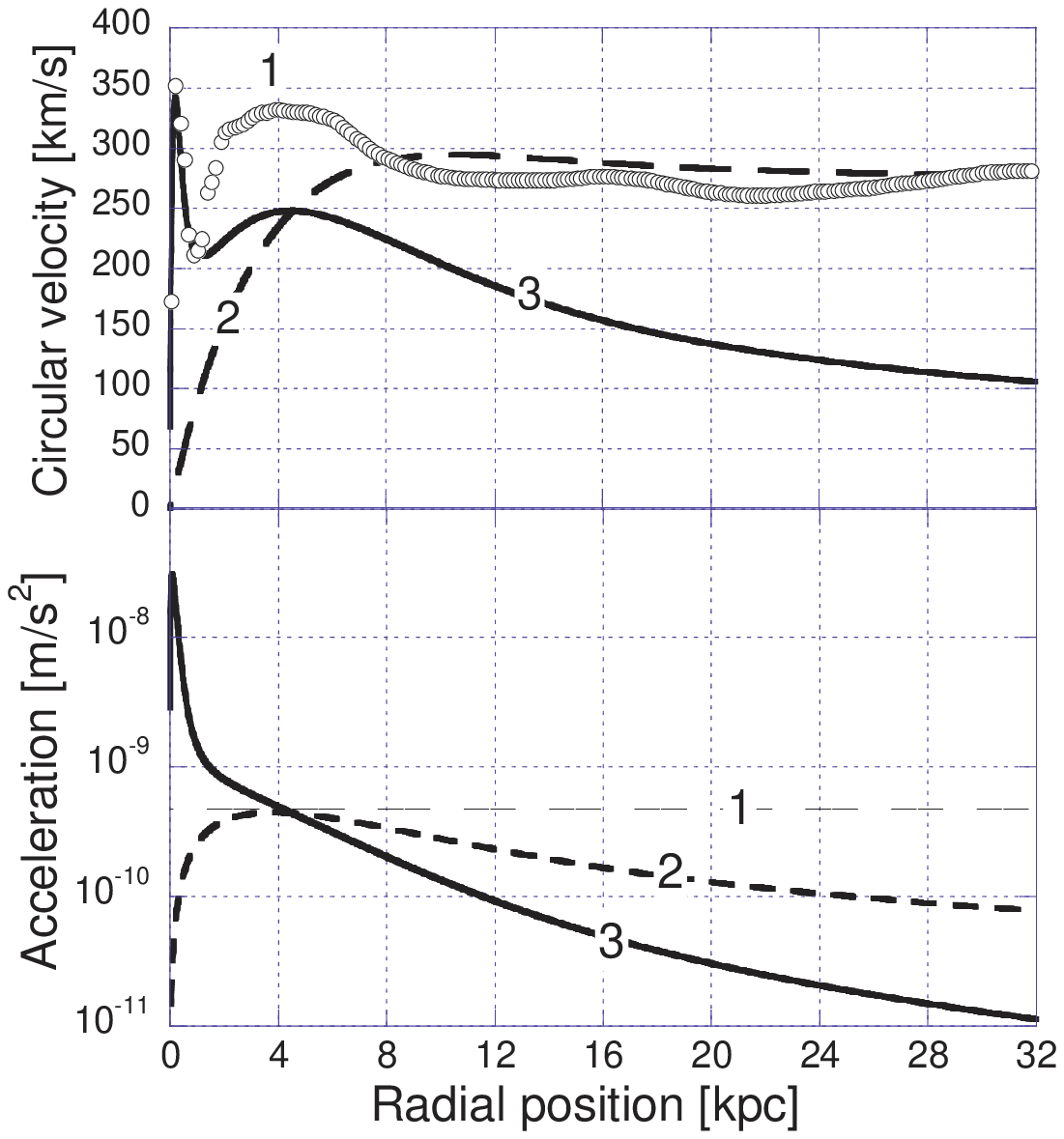}
\caption{\label{fig:fig6} (a) Above: The rotational speeds in NGC 2841 (Sb spiral) (circles - curve 1). Theoretical curve fits from the Carmelian equation (\ref{eqn:rotcurve}) (curve 2) and from the Newtonian equation (\ref{eqn:Newtonian2}) (curve 3) (b) Below: The critical acceleration $\frac{2}{3}a_{0}$ (curve 1). The corresponding rotational accelerations determined from the Carmelian (curve 2) and the Newtonian (curve 3) equations}
\end{figure}
\begin{figure}
\includegraphics[width = 3.5 in]{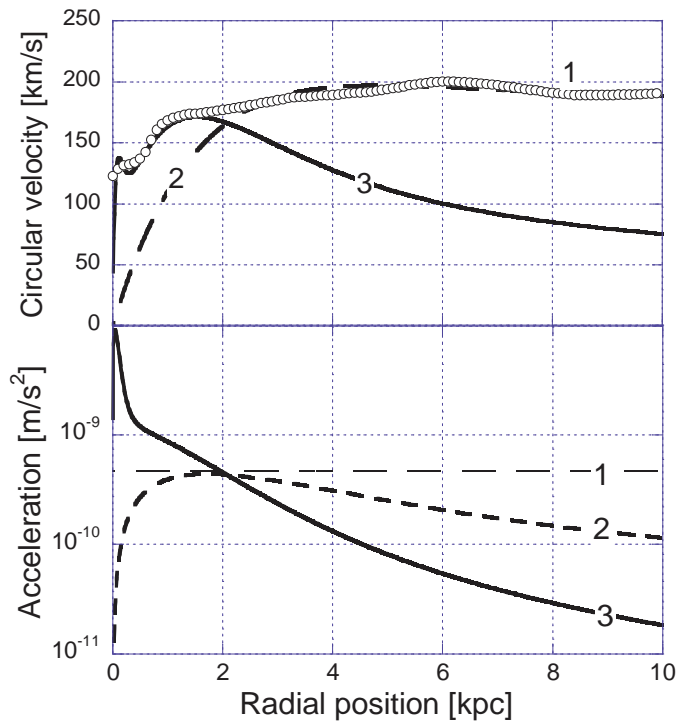}
\caption{\label{fig:fig7} (a) Above: The rotational speeds in IC 0342 (Sc spiral)((circles - curve 1). Theoretical curve fits from the Carmelian equation (\ref{eqn:rotcurve}) (curve 2) and from the Newtonian equation (\ref{eqn:Newtonian2}) (curve 3) (b) Below: The critical acceleration $\frac{2}{3}a_{0}$ (curve 1). The corresponding rotational accelerations determined from the Carmelian (curve 2) and the Newtonian (curve 3) equations}
\end{figure}

Carmeli \cite{Carmeli1998, Carmeli2000} determined a Tully-Fisher type relation using the Newtonian circular velocity equation expressed in spherical coordinates,
\begin{equation} \label{eqn:Newtonianspherical}
v^{2} = \frac{G M }{r},
\end{equation}
where it is assumed that test particles orbit at radius $r$ outside of a fixed mass $M$. Then by eliminating $r$ between (\ref{eqn:Newtonianspherical}) and (\ref{eqn:Carmelieqngoodapprox}) we get the result. This is achieved by taking the $3/2$ power of (\ref{eqn:Newtonianspherical}) and multiplying it with (\ref{eqn:Carmelieqngoodapprox}) yielding
\begin{equation} \label{eqn:TF}
v^{4} = G M \frac{2}{3}a_{0}.
\end{equation}

So by applying the same approach with (\ref{eqn:Newtonian2}) (raising it to the $3/2$ power) and multiplying it with (\ref{eqn:Carmelieqngoodapprox}) we can derive an equation describing the rotation curves in galaxies. The result is 
\begin{equation} \label{eqn:rotcurve}
v^{4} = G M \frac{2}{3}a_{0} \left\{\left(\frac{r}{2a} \right)^{9/2} 8 \; \Pi^{3/2}\right\},
\end{equation}
remembering $\Pi$ is a function of $r/2a$. It is is easily confirmed that as $ r \rightarrow \infty$,
\begin{displaymath} 
\left(\frac{r}{2a} \right)^{9/2} 8 \; \Pi^{3/2} \rightarrow 1,
\end{displaymath}
which is the radial position (or $r$) dependent part of (\ref{eqn:rotcurve}). Hence (\ref{eqn:rotcurve}) then recovers the form of the Tully-Fisher relation (\ref{eqn:TF}).

By taking the 4th root of  (\ref{eqn:rotcurve}) we get an expression for the circular velocity of test particles as a function of their radial position $r$. That result has been plotted in curve 2 of figs \ref{fig:fig2}(a) - \ref{fig:fig10}(a) for each galaxy with  $a$ and $M$ determined as fit parameters. The resulting curves have the characteristic flat shape for large radial position $r$. At small values of $r$ the rotation speeds determined from the Newtonian equation (curve 3) dominate as seen in figs \ref{fig:fig4}(a) - \ref{fig:fig10}(a). 

\section{\label{sec:Accelerations}Accelerations}

\begin{figure}
\includegraphics[width = 3.5 in]{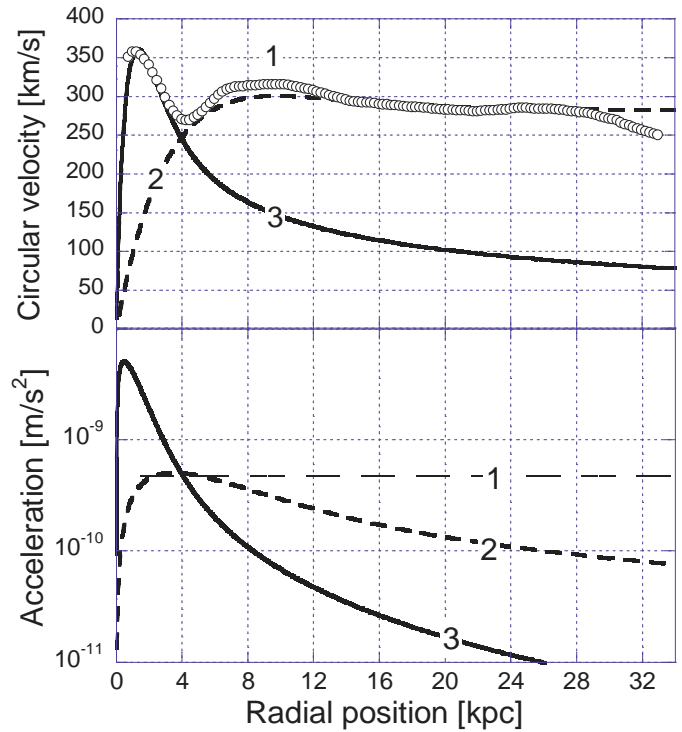}
\caption{\label{fig:fig8} (a) Above: The rotational speeds of tracer gases in NGC 1097 (SBb barbed spiral) (circles - curve 1). Theoretical curve fits from the Carmelian equation (\ref{eqn:rotcurve}) (curve 2) and from the Newtonian equation (\ref{eqn:Newtonian2}) (curve 3) (b) Below: The critical acceleration $\frac{2}{3}a_{0}$ (curve 1). The corresponding rotational accelerations determined from the Carmelian (curve 2) and the Newtonian (curve 3) equations}
\end{figure}

\begin{figure}
\includegraphics[width = 3.5 in]{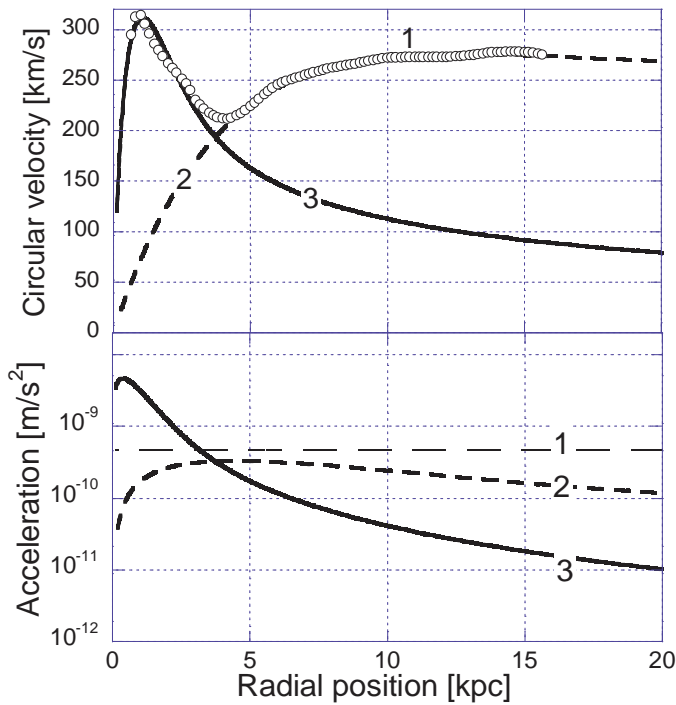}
\caption{\label{fig:fig9} (a) Above: The rotational speeds of tracer gases in NGC 2590 (Sb spiral) (circles - curve 1). Theoretical curve fits from the Carmelian equation (\ref{eqn:rotcurve}) (curve 2) and from the Newtonian equation (\ref{eqn:Newtonian2}) (curve 3)  (b) Below: The critical acceleration $\frac{2}{3}a_{0}$ (curve 1). The corresponding rotational accelerations determined from the Carmelian (curve 2) and the Newtonian (curve 3) equations}
\end{figure}

The acceleration $\frac{2}{3} a_{0}$ in (\ref{eqn:Carmelieqngoodapprox}) can be considered to be a critical acceleration. Therefore when we compare the accelerations derived from the Newtonian equation (\ref{eqn:Newtonian2}) and the Carmelian equation (\ref{eqn:rotcurve}) with this critical acceleration we notice two regimes develop. See figs \ref{fig:fig2}(b) - \ref{fig:fig10}(b). 

For example,  fig \ref{fig:fig4}(b) is very instructive. There the straight line (curve 1) is the critical acceleration $\frac{2}{3} a_{0}$, curve 2 represents the acceleration derived from the Carmelian equation (\ref{eqn:rotcurve}) and curve 3 represents the acceleration derived from the Newtonian equation (\ref{eqn:Newtonian2}). For the values, determined from the fits, of the mass ($M$) and the radial scale length ($a$), which determine how the matter density varies as a function of radial position $r$, the curves 2 and 3 cross each other very close to the critical acceleration. The significance is that for accelerations less than the critical acceleration the Carmelian force applies and for accelerations greater than the critical acceleration the Newtonian force applies. Note also that the Newtonian curve 3 has a $r^{-2}$ dependence and the Carmelian curve 2 has a $r^{-1}$ dependence above $10\, kpc$. When curves of the form $r^{-x}$ were fitted to the functions used in fig \ref{fig:fig4}(b) between 15 and 20 $kpc$, the coefficients $x$ were determined to be $x = 2.003$ and $x = 1.025$ respectively. 

From (\ref{eqn:Newtonian2}) the gravitational acceleration ($v^{2}/r$) can be calculated in the limit of $r \rightarrow \infty$, outside most of the matter of the galaxy. As expected for the Newtonian model it tends to $GM/r^{2}$. And similarly from (\ref{eqn:rotcurve}) the gravitational acceleration ($v^{2}/r$) can also be calculated in the limit of large $r$, for the Carmelian model. In this case it is evident from (\ref{eqn:TF}) that it must tend to $\sqrt{G M \frac{2}{3}a_{0}}/r$. In this regime the accelerations are very weak. This is very significant as alternative theories of gravity have been suggested (for example, \cite{Milgrom1983a,Milgrom1983b,Milgrom1983c}) where the force of gravity falls away as $r^{-1}$ for small accelerations. However for small $r$, that is, where $r \rightarrow 0$, close to the origin of the central gravitational potential, the effect of the Carmelian force law becomes extremely small and is many orders of magnitude smaller than that for the Newtonian force law.


\section{\label{sec:Galaxies}Sample of Galaxy Rotation Curves}

\begin{figure}
\includegraphics[width = 3.5 in]{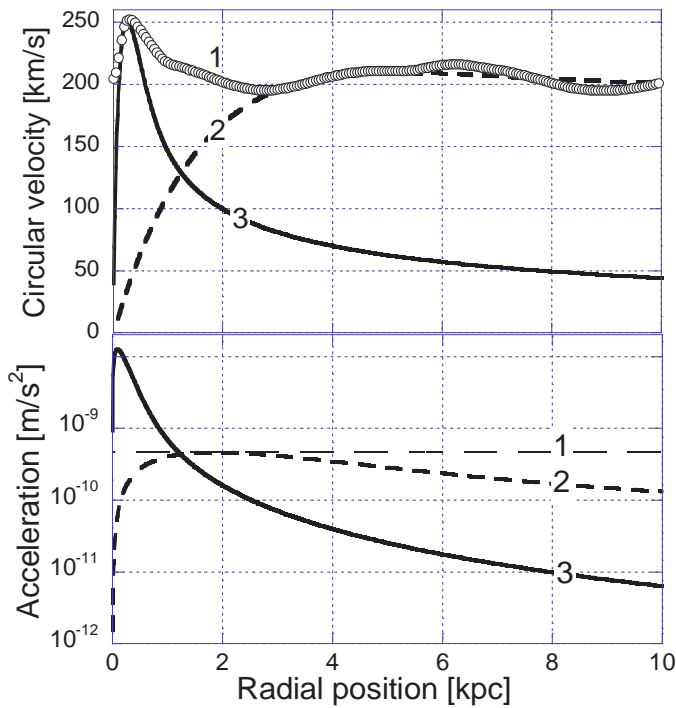}
\caption{\label{fig:fig10} (a) Above: The rotational speeds of tracer gases in the Galaxy (Milky Way Sb spiral). Measured data from \cite{Sofue1999} shown in curve 1. Theoretical curve from the Carmelian equation (\ref{eqn:rotcurve}) (curve 2) and from the Newtonian equation (\ref{eqn:Newtonian2}) (curve 3) (b) Below: The critical acceleration $\frac{2}{3}a_{0}$ (curve 1). The corresponding rotational accelerations determined from the Carmelian (curve 2) and the Newtonian (curve 3) equations respectively}
\end{figure}

A sample of 9 galaxy fits are shown in figs \ref{fig:fig2} - \ref{fig:fig10}. The top (a) figures show the rotation curve fits and the bottom (b) figures show the resulting acceleration regimes. In each figure, the measured rotational speeds of tracer gases in the chosen spiral galaxy is shown as as a function of radial position (curve 1). Measured data are taken from \cite{Sofue1999}. Theoretical curves from the Carmelian equation (\ref{eqn:rotcurve}) (curve 2) and from the Newtonian equation (\ref{eqn:Newtonian2}) (curve 3) are fitted over the range of $r$ which best fit the data by allowing $a$ and $M$ to be free parameters. The accelerations in the bottom (b) figures are derived from the Carmelian (curve 2) and the Newtonian (curve 3) equations respectively with values of $a$ and $M$ derived from the fits in the (a) figures. These are compared with (curve 1) the critical acceleration $\frac{2}{3}a_{0} \approx 4.75 \times 10^{-10}$ $m.s^{-2}$ determined elsewhere from $\tau \approx 4.21 \times 10^{17} s$, which is the reciprocal of the Hubble parameter at zero distance $h \approx 73.270$ \kmsmpc.

\subsection{\label{sec:Galaxy}Extragalactic spirals}

In this section, I discuss individual galaxy curve fits, starting with fig \ref{fig:fig2} showing the barbed spiral NGC 3198. In each case, because of the possibility of different acceleration regimes, both Carmelian and Newtonian curve fit were attempted. In fig \ref{fig:fig2}(a) the Carmelian fit is shown by the broken curve 2 to be the only good fit. The scale radius $a = 1.85\; kpc$ and $M=0.984 \times 10^{10} M_{\odot}$ determined from the fit. Curve 3 shows the best Newtonian fit with $a = 2.99\; kpc$ and $M=4.2 \times 10^{10} M_{\odot}$ but it doesn't fit well at high values of $r$. The Newtonian fit results in a mass at least 4 times greater than that from the Carmelian fit. The scale radius determined from luminous matter for this galaxy is $a=2.5 \; kpc$ which is closer to the Carmelian curve determination.

In the fig \ref{fig:fig2}(b) curve 2 shows the acceleration using the Carmelian  determined values of the scale radius $a$ and mass $M$. Curve 3 shows the acceleration for Newtonian fit determined values. Clearly curve 2 is dominant and is always less than the critical acceleration $\frac{2}{3}a_{0}$. In this model, when the accelerations are less than the critical value the Carmelain force applies.

Next the data for NGC 0598, a Sc spiral galaxy, shown in fig \ref{fig:fig3}(a), fits both a Carmelian and a Newtonian curve. From the Carmelian model the scale radius $a = 1.85\; kpc$ and $M=0.13 \times 10^{10} M_{\odot}$ determined from the fit. Clearly the Carmelian curve 2 is the better fit over the Newtonian curve 3 for the following reasons. Firstly the data (circles - curve 1) continue to rise or at least are not falling at the extremity of the available measured range. The Newtonian curve indicates it should fall. Secondly from fig \ref{fig:fig3}(b) the accelerations are much less than the critical value $\frac{2}{3}a_{0}$ and hence in this regime the Carmelian force law applies.

From the Newtonian model a scale radius of $a=2.22\;kpc$ and $M=1.42 \times 10^{10} M_{\odot}$ were determined but the fit doesn't conform to the model. Nevertheless the Newtonian fit results in a mass 10 times greater than the fit for the Carmelian model.

Figs \ref{fig:fig4} and \ref{fig:fig5} show both Newtonian and Carmelian models fit the rotation curve data for the galaxies NGC 2903 and NGC 7331. In the high acceleration regime a Newtonian fit is applicable and when the acceleration drops below $\frac{2}{3}a_{0}$ the Carmelian applies. From the Table the best fit determined values of $a$ and $M$ for each galaxy are listed. They are compared with published values of $a$ and masses determined from different methods. 

In these and all subsequent top (a) figures, the fits for curves 2 and 3, respectively, apply only for accelerations less than and greater than the critical acceleration. In the rotation curve fits, in the top (a) figures, the circular velocities are not added but each apply over their respective regimes. This means the masses determined from the Newtonian fits must be less than those from the Carmelian fits because the Newtonian determined mass must lay within the radius $R$ where the two curves cross. The Newtonian fits occur in the stronger acceleration regimes ($> \frac{2}{3}a_{0}$) close to the galactic center as indicated by all of the bottom (b) figures.

The Table lists the radial position ($R$) where the Newtonian and Carmelian regimes meet, the mass ($M_{R}$) out to a radius $r = R$  and the total mass $M$ shown in Col. 8 (Carmeli $M$ from fit), which is the Carmelian model determined mass. As expected the values of $M_{R} \leq M$ determined from the Newtonian regime.

Figs \ref{fig:fig6} and \ref{fig:fig7}, respectively, show NGC 2841 and IC 0342, which have been modeled with two central components with different scale radii.  Both seem to have a dense mass concentration toward their centers. The scale radii for these inner most concentrations are $a = 0.09 \, kpc$ and $a = 0.05 \,kpc$ for NGC 2841 and IC 0342 respectively.  The combined rotation curve for IC 0342 is a much better fit to the data using the two-acceleration-regime model than that of the former, NGC 2841. Nevertheless the theory works well in these type of galaxies also. 

Figs \ref{fig:fig8} and \ref{fig:fig9}, respectively, show NGC 1097 and NGC 2590 modeled with only one central dense Newtonian component and they fit the model very well. Deviations in most cases I believe can be attributed to the fact that no azimuthal dependence has been added to the model, nor does it allow for anything but constant scale radii over  regions of the galaxy where the fits apply. That is clearly unphysical but seems to be a reasonable approximation.  

The Table also shows, in Col. 9, the mass ($M_{10}$) determined at $r$ = 10 $kpc$ using the Newtonian formula
\begin{equation} \label{eqn:Keplermass}
M = \frac{v^2 r}{G}
\end{equation}
assuming that most of the mass is internal to $r =10\;kpc$ and the measured rotation speeds. This calculation is compared with the mass ($M_{10}$), in Col. 10, derived from the Carmelian equation with $r$ = 10 $kpc$. The ratios of these masses are shown in Col. 11. It indicates that using normal Newtonian/Keplerian dynamics seriously over estimates galaxy masses by between 2 and 7 times. These values are consistent with the needed mass levels of halo dark matter to achieve the correct rotation curves.

\begin{table*}[t]
\small
\begin{center}
Table:~ \hspace{4pt} Important galaxy data where valid curve fits are found
\end{center}
\vspace{6pt}
\begin{tabular*}{\textwidth}{@{\hspace{\tabcolsep}
\extracolsep{\fill}}lccccccccccccc}
\hline\hline
   & & &Scale radius &Newton &Carmeli &Newton &Carmeli &Newton &Carmeli & &Radius \\
  Figure &Galaxy &Type &$a$ &$a$ &$a$ &$M$ &$M$ &$M_{10}$ &$M_{10}$ &Ratio &$R$ &$M_{R}$\\
  		&Name &  &published &from fit &from fit  &from fit  &from fit & & &\\
\hline
2 & NGC 3198 & SBc & 2.5 & -    & 1.85 & -    & 0.984 & 6.554  & 0.956 & 6.85 & -    & - \\
3 & NGC 0598 & Sc  & 2.7 & -    & 1.01 & -    & 0.13  & -      & -     & -    & -    & - \\
4 & NGC 2903 & Sc  & 1.9 & 0.31 & 0.98 & 0.54 & 2.12  & 8.810  & 2.120 & 4.16 & 1.39 & 0.54 \\
5 & NGC 7331 & Sbc & 4.7 & 1.15 & 1.2  & 4.0  & 4.45  & 11.496 & 4.440 & 2.59 & 3.38 & 3.16 \\
6 & NGC 2841 & Sb  & 2.3 & 2.26 & 2.30 & 8.12 & 9.00  & 17.721 & 8.377 & 2.11 & 2.11 & 5.09 \\
7 & IC 0342  & Sc  &     & 0.74 & 1.05 & 1.28 & 1.81  & 8.200  & 1.809 & 4.53 & 2.08 & 1.00 \\
8 & NGC 1097 & SBb &     & 0.62 & 2.12 & 4.80 & 9.74  & 22.997 & 9.242 & 2.49 & 3.94 & 4.74\\
9 & NGC 2590 & Sb  & 2.1 & 0.5  & 2.73 & 2.90 & 6.96  & 17.213 & 6.127 & 2.81 & 3.73 & 2.89\\
10 & MW Galaxy & Sb &    & 0.12 & 1.09 & 0.45 & 2.31  & 9.302  & 2.308 & 4.03 & 1.22 & 0.45\\
\hline
\end{tabular*}
\vskip 2mm
\noindent Cols (4) -(6): $a$ in $kpc$.
Cols (7)-(10), (13): $M$ in [$10^{10} M_{\odot}$].
Cols (9) $\&$ (10): Mass ($M_{10}$) calculated at $r=10\;kpc$.
Col.(11): Mass from Col. (9) divided by Col. (10).
Cols (12): $R$ in $kpc$.
Col. (13): Mass ($M_{R}$) calculated at $r = R$.
\end{table*}

\subsection{\label{sec:MWGalaxy}The Galaxy}

Considering the (Milky Way) Galaxy the same analysis has been applied to the data from \cite{Honma1997}. See the rotation curve in fig \ref{fig:fig10}(a). Other observers don't record the central peak indicative of a large central mass concentration. However the compact radio source Sagittarius A* at the Galactic center is widely believed to be associated with the supermassive black hole with a mass of about $3.59 \pm 0.59 \times 10^6 M_{\odot}$. \cite{Eisenhauer2003} Then there is the vast concentration of matter in the Galactic bulge. From the Newtonian curve fit we'd expect that within 1.2 $kpc$ of the center there is a mass of about $M=4.5 \times 10^{9} M_{\odot}$. 

The Carmelian curve fit (curve 2 in fig \ref{fig:fig10}(a)) over the range 3 to 10 $kpc$ is an excellent fit. The acceleration regimes, consistent with the curve fits (curves 2 and 3) in fig \ref{fig:fig10}(a), are shown in fig \ref{fig:fig10}(b). Except for the intervening region the fits agree well with the theory. The discrepancy could be due to an unmodeled higher concentration of mass in the central bulge region. Deviations in the spiral arms from a smooth exponential density decay are consistent with the oscillations about curve 2 in fig \ref{fig:fig10}(a).

The distance of our solar system from the Galactic center has been estimated at $9.95 \; kpc$ \cite{Honma1997} and more recently as $7.94 \pm 0.42 \; kpc$ \cite{Eisenhauer2003}. Likewise the enclosed mass may be calculated using both (\ref{eqn:Keplermass}) and (\ref{eqn:rotcurve}). When one uses the normal Newtonian/Keplerian calculation (\ref{eqn:Keplermass}) and $v = 200.78\, km.s^{-1}$ the speed of the solar system orbiting the Galactic center, it results in an estimate of the enclosed mass of $M_{10} =9.3 \times 10^{10} M_{\odot}$ as compared with $M_{10}=2.3 \times 10^{10} M_{\odot}$ from the Carmelian calculation (\ref{eqn:rotcurve}), which is 4 times smaller. For a distance $7.94 \; kpc$ the Keplerian calculation of the enclosed mass is $M_{10} =7.6 \times 10^{10} M_{\odot}$, which is 3.3 times greater than the Carmelian calculation for the enclosed mass at that distance.

\section{\label{sec:Conclusion}Conclusion}
Carmeli's Cosmological General Relativity theory provides a solution to the rotation curve anomaly in the outer regions of spiral galaxies. Equations of motion have been derived from Carmeli's metric assuming a gravitational potential in cylindrical coordinates. A Tully-Fisher type relation results and the rotation curves for spiral galaxies are accurately reproduced without the need for non-baryonic halo dark matter. 

Two acceleration regimes are discovered that are separated by a critical acceleration $\frac{2}{3} a_{0}$. For accelerations larger than the critical value the Newtonian force law applies, but for accelerations less than the critical value the Carmelian regime applies. In the Newtonian regime the accelerations fall off as $r^{-2}$ as expected, but in the Carmelian regime the accelerations fall off as $r^{-1}$. This is new physics but is exactly what is suggested by Milgrom's phenomenological MOND theory.  However in this case a theoretical basis is found, whereas until now, no theory has been found for Milgrom's MOND.  This theory also provides an understanding of the connection between the two regimes.
\\
\section{Acknowledgment}
I would like to thank Prof. Sofue for supplying me with electronic rotation curve data and Nick Loh for valuable discussion about cylindrical coordinates.\\

\end{document}